\documentclass[prd,aps,showpacs,showkeys]{revtex4-1}
\usepackage{bm,amsfonts,latexsym,amsmath,amssymb,amsbsy,graphicx}

\newcommand{\bb}{\begin{eqnarray}}
\newcommand{\ee}{\end{eqnarray}}
\newcommand{\ba}{\begin{align}}
\newcommand{\ea}{\end{align}}

\begin{document}

\title{\bf Quasi-stationary states and fermion pair creation from a vacuum in supercritical Coulomb field}
\author{V.R. Khalilov}\email{khalilov@phys.msu.ru}
\affiliation{Faculty of Physics, M.V. Lomonosov Moscow State University, 119991,
Moscow, Russia}

\begin{abstract}

Creation of charged fermion pair from a vacuum in the so-called supercritical Coulomb potential
is examined for the case when created pair moves in one plane. In which case
the quantum dynamics of charged massive or massless fermions can be
described by the two-dimensional Dirac Hamiltonians with a Coulomb
potential. These Hamiltonians are singular and require the additional
definition in order for them to be treated as  self-adjoint quantum-mechanical operators.
We construct the self-adjoint  two-dimensional Dirac Hamiltonians with a Coulomb
potential and determine the quantum-mechanical states for such Hamiltonians
in the corresponding Hilbert spaces of square-integrable functions.
We determine the scattering amplitude in which  the self-adjoint extension
parameter is incorporated and then obtain the equations implicitly defining
the possible discrete energy spectra of the self-adjoint Dirac Hamiltonians
with a Coulomb potential. It is shown that the quantum system in the presence of
a supercritical Coulomb potential becomes unstable which manifests
in the appearance of quasi-stationary states  in the lower (negative) energy continuum.
The energy spectrum of these states is quasi-discrete, consists of
broadened levels whose width is related to the inverse lifetime
of the  quasi-stationary state as well as the creation probability of charged fermion pair
by supercritical Coulomb field. 
Explicit analytical expressions for the creation probabilities
 of charged (massive or massless) fermion pair  are obtained in a supercritical Coulomb field.
\end{abstract}

\pacs{12.20.-m, 03.65.Ge, 71.55.-i}

\keywords{Supercritical Coulomb potential; Vacuum instability; Scattering amplitude poles; Quasi-stationary states; Fermion pair creation}

\maketitle

\section{Introduction}

The instability of quantum electrodynamics vacuum  in the presence
of the so-called supercritical Coulomb potential  of a hypothetical atomic nucleus
with the charge of a nucleus $Z$ exceeding a certain critical value $Z_{cr}\sim 170$
($Z_{cr}\alpha\sim 1.24$, $\alpha\approx 1/137$ is the fine structure constant)
have been studied a long time \cite{003,004,blp,grrein,01,02,03,04}.
It has been understood  this phenomenon is related to electron-positron
pair creation from a vacuum. Because of the absence of such
supercharged  atomic nuclei the electron-positron pair creation
due to such an instability  is highly academic problem.
Nevertheless, significant efforts to observe the positrons created due to
the vacuum instability have been made through colliding heavy-ions with enormously large
$Z\sim 170$ but results  look ambiguous.

The vacuum instability was found to occur in  the spatially two-dimensional
quantum systems in which case the Coulomb potential strength
$Z_{cr}\alpha \sim 0.65$ ($Z_{cr}\sim 90$) for massive case \cite{khho}
and $Z_{cr}\alpha_g =0.5$ ($\alpha_g$ is the the effective coupling constant)
for massless case \cite{vp11a,as11b,as112,mmf,vmvn,ggg,wzq}.
So the interest in similar phenomena in two-dimensional relativistic
quantum systems was revived in connection with the Coulomb impurity problem in
graphene \cite{vp11a,as11b,as112,6,7,8,review,10,11,15}. Indeed,
 charge carriers in  graphene behave as relativistic particles
described by the two-dimensional Dirac equation \cite{vp11a,as11b,10,11,15},
which allows to consider  graphene as the condensed matter analog
of the quantum electrodynamics in 2+1 dimensions \cite{datdc,ggvo}.
Besides, in graphene,  the corresponding ``effective fine structure constant"
$\alpha_{eff}=e^2/\epsilon_0\hbar v_F$ (where $e$ is the electron charge,
$\epsilon_0$ is the dielectric constant of the medium, $\hbar$ is the action constant
and $v_F$ is the Fermi-Dirac velocity) is large ($\alpha_{eff}\sim 1$) \cite{7,10,13}
and a cluster of charged impurities can produce the supercritical Coulomb potential,
which opens the real possibility of testing the supercritical instability \cite{review}.
The electron-hole pair creation (holes in graphene play the role
of positrons \cite{vp11a,as11b,as112}) is likely to be now revealed in graphene (see, \cite{gr0,gr1,gr2}).

The induced electric current  due to vacuum polarization of massless fermions
in the superposition of Coulomb and Aharonov--Bohm potentials in 2+1 dimensions was addressed
in  \cite{khim}. Vacuum polarization of the massive charged fermions
can also be of interest for graphene with Coulomb impurity  \cite{gkg1}; the plane density of an induced vacuum charge in a strong  Coulomb potential for massless and massive
fermions was studied in \cite{khim1}.

In present paper we study the creation of charged fermion pair from a vacuum by
supercritical Coulomb potential for the case when the quantum dynamics of the massive
or massless fermions are governed by the two-dimensional Dirac Hamiltonians
with a Coulomb potential. These Dirac Hamiltonians are singular and require the supplementary
definition in order for them to be treated as  self-adjoint quantum-mechanical operators.
Self-adjoint  Hamiltonians are not unique but each of them
can be specified a real "self-adjoint extension" parameter
by additional (self-adjoint) boundary conditions. To put it more exactly,
a domain, including the singular $r = 0$ region, in the Hilbert
space of square-integrable functions must be indicated for each self-adjoint Hamiltonian.

We find  the quantum-mechanical states for  self-adjoint two-dimensional
Hamiltonians with a Coulomb potential by constructing the corresponding Hilbert spaces.
We determine the scattering amplitude in which  the self-adjoint extension
parameter is incorporated and then obtain the equations implicitly defining
the energy spectra of these Hamiltonians. It is shown that the quantum system
in  the presence of supercritical Coulomb potential 
becomes unstable which manifests in the appearance of quasi-stationary states 
in the lower (negative) energy continuum. 

The scattering amplitude in a supercritical  Coulomb potential becomes ambiguous function;
it has  a discontinuity in the complex plane of energy and additional poles
on the negative energy axis of the second (nonphysical) sheet of the Riemann surface. 
The quasi-discrete spectrum of quasi-stationary  states consists of broadened levels 
whose width (defined by the imaginary part of energy $E$) is related to
the inverse lifetime (the decay rate) of the  quasi-stationary state.
We derive equations for the energy spectra and quasi-stationary state lifetimes
and analyze (solve) these equations in physically important cases.
The quasi-stationary states  are directly associated with the fermion pair creation
in the quantum electrodynamics and the modulus of the imaginary part $E$ determines the doubled probability of the creation of charged fermion pair by Coulomb potential.

We shall adopt the units where $c=\hbar=1$.

\section{Spectra of the self-adjoint radial Dirac Hamiltonians}
We are only interested on the planar dynamics of a charged fermion
in a Coulomb potential what implies that the fermion moves in the $xy$ plane and the
fermion momentum projection $p_z = 0$,  together with the imposition of the Coulomb field should
be now intrinsically two-dimensional (see, for instance, \cite{sil}).
For which reason we shall assume in what follows that the space of fermion quantum
states is the two-dimensional Hilbert space $\mathfrak H=L^2(\mathbb R^2)$ of square-integrable
functions $\Psi({\bf r}), {\bf r}=(x,y)$ with the scalar product
\bb
(\Psi_1,\Psi_2)=\int \Psi_1^{\dagger}({\bf r})\Psi_2({\bf r})d{\bf r},\quad d{\bf r}=dxdy.
\label{scpr}
\ee
One also knows \cite{27} that  the two-dimensional Dirac matrices can be represented  in terms
of the Pauli matrices, namely,
 $\gamma^0= \sigma_3,\quad \gamma^1=is\sigma_1,\quad \gamma^2=i\sigma_2$, where  the parameter $s=\pm 1$ can be introduced to label two types of fermions in accordance with the
signature of the two-dimensional Dirac matrices \cite{27}; for the case of massive fermions it can be applied to characterize two states of the fermion spin (spin "up" and "down")  \cite{4}.

Then, the Dirac Hamiltonian for a fermion of the mass $m$ and charge
$e=-e_0<0$ in a Coulomb potential $A_0(r) =a/e_0r$, $A_r=0$, $A_{\varphi}=0$, $a>0$
in polar coordinates $r=\sqrt{x^2+y^2}$, $\varphi=\arctan(y/x)$, is
\bb
 H_D=\sigma_1P_2-s\sigma_2P_1+\sigma_3 m-e_0A_0(r),\label{diham}
\ee
where $P_\mu = -i\partial_{\mu} - eA_{\mu}$ is the
generalized fermion momentum operator (a three-vector).
The Hamiltonian (\ref{diham}) should
be defined as a self-adjoint operator in the Hilbert space
of square-integrable doublets $\Psi({\bf r}), {\bf r}=(x,y)$
with the scalar product (\ref{scpr}).

Eigenfunctions of the Hamiltonian (\ref{diham}) can be represented as (see, for instance \cite{khpr,khle1,khle0})
\bb
 \Psi(t,{\bf r}) = \frac{1}{\sqrt{2\pi r}}
\left( \begin{array}{c}
f_1(r)\\
f_2(r)e^{is\varphi}
\end{array}\right)\exp(-iEt+il\varphi)~, \label{three}
\ee
where $E$ is the fermion energy, $l$ is an integer.
The wave function $\Psi$ is an eigenfunction of the
total angular momentum $J\equiv L_z+ s\sigma_3/2$, where $L_z\equiv
-i\partial/\partial\varphi$, with eigenvalue $j=l+s/2$ and
\bb h F= EF, \quad F=\left(
\begin{array}{c}
f_1(r)\\
f_2(r)\end{array}\right), \label{radh}\ee
where
\bb
 h=is\sigma_2\frac{d}{dr}+\sigma_1\frac{l+\mu+s/2}{r}+\sigma_3m-\frac{a}{r}.
\label{radh0}
\ee

The radial Hamiltonian $h$ is  singular and  so the supplementary
definition is required in order for it to be treated as
a self-adjoint quantum-mechanical operator, therefore, we must indicate
the Hamiltonian domain in the Hilbert space of square-integrable functions
on the half-line, including the $r = 0$ region.
One knows that self-adjoint Dirac Hamiltonians can be constructed
for a symmetrical operator $h$, if
\bb
 \int\limits_{0}^{\infty} G^{\dagger}(r)h F(r) dr =
 \int\limits_{0}^{\infty}[h G(r)]^{\dagger}F(r) dr  \label{sym}
\ee
for any doublets $F(r)$ and $G(r)$.

Let us define the operator $h^0$ in the Hilbert space $\mathfrak L^2(0,\infty)$ as
$$
h^0{:}\left\{\begin{array}{l}
 D(h^0)=D(0,\infty), \\
 h^0F(r)= h F(r),
\end{array}\right.
$$
where  $D(0,\infty)$ is the standard space of
smooth functions on $(0,\infty)$ vanishing at $r\to\infty$.
It is evident that $h^0$ is the symmetrical operator.

Let $h$ be the self-adjoint extension of $h^0$ in $\mathfrak L^2(0,\infty)$ and
let us consider the adjoint operator  $h^*$ given by Eq. (\ref{radh0}) but
defined as follows
\bb
h^*{:}\left\{\begin{array}{l}
 D(h^*)=\left\{\begin{array}{l}
F(r):\;F(r)\;\mbox{are absolutely continuous in}(0,\infty),\\
F, h F \in{\mathfrak L}^2(0,\infty),
\end{array}\right.\\
h^*F(r)=hF(r),
\label{hadj}
\end{array}\right.
\ee
i.e. $D(h^0)\subset D(h^*)$.
A symmetric operator $h$ is self-adjoint, if its domain $D(h)$
coincides with that of its adjoint operator  $D(h^*)\equiv D^*$.

As far as $\lim\limits_{r\to \infty} F(r)=0$ for any $F(r)$ of $D(h^*$, integrating (\ref{sym}) by parts, we reduce Eq. (\ref{sym}) to boundary conditions at $r=0$
\bb
\lim_{r\to 0} G^{\dagger}(r)i\sigma_2 F(r)=0. \label{bounsym}
\ee

If (\ref{bounsym}) is satisfied for any doublet $F(r)$ of $D^*$
then the operator  $h^*$ is symmetric and, so, self-adjoint.
This means that  the operator $h$ is essentially self-adjoint, i.e.,
its unique self-adjoint extension is its closure $h= \check h$,
which coincides with the adjoint operator $h=h^*$.
If (\ref{bounsym}) is not satisfied then the self-adjoint operator
$h=h^{\dagger}$ can be found as the narrowing  of  $h^*$ on the so-called
maximum domain $D(h)\subset D^*$.

Thus, any doublet $F(r, E)$ of $D^*$ must satisfy boundary condition \cite{vgt}
\bb
 (F^{\dagger}(r, E)i\sigma_2 F(r, E))|_{r=0}= (f_1^*f_2-f_2^*f_1)|_{r=0} =0, \label{bouncon}
\ee
where $f^*$ is the complex conjugate function.
Physically, Eq. (\ref{bouncon}) shows that  the probability current density
is equal to zero at the origin.

The needed regular (at $r=0$) doublet can be conveniently represented as
\bb
F(r,E)=
\left(
\begin{array}{c}
f_1(r)\\
f_2(r)\end{array}\right)=Ar^{\gamma}e^{ipr}\left(\begin{array}{c}
	\sqrt{E+m} \\ -i\sqrt{E-m}
\end{array}\right)\left[\Phi(b,c;x)\pm \frac{\gamma-iaE/p}{\nu-iam/p}\Phi(b+1,c;x)\right].
\label{gensol}
\ee
Here
\bb
\gamma=\sqrt{(l+1/2)^2-a^2}, \nu=l+1/2, b=\gamma-iaE/p,c=2\gamma+1, x=-2ipr, p=\sqrt{E^2-m^2},
\label{not}
\ee
$A$ is a  constant, $\Phi(b, c; x)$ is the confluent hypergeometric function \cite{GR} and because the fermion states in a Coulomb potential are doubly degenerate with respect
to $s$ we put $s=1$.

The asymptotic behavior of doublets (wave functions)
at $r\to 0$ is determined with quantity  $\gamma$,
which is real for $a^2\leq \nu^2$ and is imaginary
$\gamma=i\sqrt{a^2-\nu^2}\equiv i\sigma$ for $a^2>\nu^2$.
Applying known formula
$$
\Phi(b, c; x)=e^x\Phi(c-b, c;-x),
$$
we have for $\Phi(b+1, c; x)$
\bb
\Phi(b+1, c; x)=e^x\Phi^*(b, c; x),
\label{hyp}
\ee
and so Eq.(\ref{gensol} takes the form
\bb
\left(
\begin{array}{c}
f_1(r)\\
f_2(r)\end{array}\right)=A'r^{\gamma}e^{-i\eta(\gamma)}\sqrt{E\pm m}\begin{array}{c}
	{\rm Re} \\ {\rm Im}
\end{array}\left[e^{ipr+i\eta(\gamma)}\Phi(b,c;x)\right]
\equiv Y(r,\gamma,E), \label{main0}
\ee
where
\bb
e^{-2i\eta(\gamma)}=\frac{\gamma-iaE/p}{\nu-iam/p}
\label{eta+}
\ee
and $A'$ is the normalization constant.

One can show that for $\gamma\ne n/2$, $n=1, 2 ,\ldots$ two doublets
\bb
U_1(r,\gamma,E)=Y(r,\gamma,E),\quad U_2(r,\gamma,E)=Y(r,-\gamma,E)
\label{1e35}
\ee
are linear independent; their asymptotic behavior
as $r\to 0$ is given by
\bb
U_1(r,\gamma,E)=(r)^{\gamma}u_+{+}O(r^{\gamma+1}),\quad
 U_2(r,\gamma,E)=(r)^{-\gamma}u_-{+}O(r^{-\gamma+1}),
\label{e36}
\ee
where
\bb
u_{\pm}=\left(\begin{array}{c}
\displaystyle u(\pm \gamma, a) \\ 1
\end{array}\right).
\ee
We see that the Hamiltonian domain in the Hilbert space of square-integrable functions is specified by the two doublets $U_{1,2}(r,\gamma,E)$ and therefore essentially depends on $\gamma$ as well as on $a$. One can separate out three regions of the values of $\gamma$.

In the region $\gamma\ge 1/2$, only the wave function $\sim U_1(r,\gamma,E)/\sqrt{r}$ is square integrable at $r=0$, but $\sim U_2(r,\gamma,E)/\sqrt{r}$ is not. For $\gamma\ge 1/2$, the wave function $\sim U_1(r,\gamma,E)/\sqrt{r}$ is regular at $r=0$. The generalized eigenfunctions $F(r,\gamma, E)$ of the radial self-adjoint  Hamiltonian are $U_1(r,\gamma,E)$. Its energy spectrum is continuous in the region  $E\geq m$ and discrete levels in the region $m>E>0$  have to exist in a Coulomb potential in addition to continuous part of the energy spectrum.

 One knows that bound states are identified as the
poles of the scattering amplitude at the analytic continuation in the region $E<m$ \cite{blp}; these poles are located on the first (physical) sheet of the Riemann surface ${\rm Re}\sqrt{m^2-E^2}>0, {\rm Im}\sqrt{m^2-E^2}>0 $  in the complex $E$-plane. The scattering amplitude relates incoming and
outgoing wave functions of a quantum system undergoing a scattering process.
In this manner, the poles are determined at the zeros
of the coefficient in the ingoing wave at the analytic continuation in the region $E<m$ (see, for instance \cite{ll}). It will be recalled that the asymptotic form of radial
doublets at $r\to \infty$ in the case under study is
\bb
\left(
\begin{array}{c}
f_1(r)\\
f_2(r)\end{array}\right)=D\sqrt{E\pm m}\begin{array}{c}
	\cos  \\ \sin
\end{array}\left(pr-\frac{\pi}{2}|l|-\frac{\pi}{4}+\delta_l\right), \label{as0}
\ee
where $D$ is a constant and the phase shifts $\delta_l$  are determined
by the potential at small $r$.

The asymptotic behavior of the wave function at $r\to \infty$ one can find by means of
the formula for the confluent hypergeometric function at $z\to \infty$
\bb
\Phi(b, c; z)=\frac{\Gamma(c)}{\Gamma(c-b)}(-z)^{-b}+ \frac{\Gamma(c)}{\Gamma(b)}e^z z^{b-c}
\label{asym0}
\ee
in which the first (leading asymptotic) term only is significant.
Here $\Gamma(z)$ is the Gamma function \cite{GR}.
After simple calculations, we obtain for the ingoing wave at $r\to \infty$
\bb
B_l(\gamma, E)\frac{e^{-i(pr-\pi|l|/2-\pi/4+ C\ln pr)}}{\sqrt{r}},
\label{ingwa}
\ee
where
\bb
B_l(\gamma, E)=\frac{\Gamma(\gamma+1+iaE/p)}{\Gamma(\gamma+1-iaE/p)}e^{-2i\eta(\gamma)-i\pi(|l+1/2|-\gamma)}
\label{coef1}
\ee
and $C=aE/p$.

Analytic continuation in the region $E<m$ on the first sheet is performing by means
of replacements
\bb
\sqrt{E-m}\to i\sqrt{m-E}, p\to i\lambda, \lambda=\sqrt{m^2-E^2}, aE/p\to -iaE/\lambda,
\label{1sheet}
\ee
and as a result, we obtain
\bb
B_l(\gamma, \lambda, E)=\frac{\Gamma(\gamma+1+aE/\lambda)}{\Gamma(\gamma+1-aE/\lambda)}\frac{\gamma-aE/\lambda} {\nu-am/\lambda}e^{-i\pi(|l+1/2|-\gamma)}.
\label{coef2}
\ee
Discrete energy levels of  bound states are defined as roots of the equation
$B_l(\gamma, \lambda, E)=0$ (as they say, the zeros of $B_l(\gamma, \lambda, E)$), i.e., either
\bb
\gamma+1-aE/\lambda=-n, \quad n=0,1,2\ldots
\label{spec0}
\ee
(at these points $\Gamma(\gamma+1-aE/\lambda)$ has the poles including a pole at $l=0,1/2>\gamma>0$)
or
\bb
\gamma-aE/\lambda=0
\label{dissp0}
\ee
(in which case $\nu<0, l=-1$).
Therefore, the discrete energy spectrum of  bound states at $a\leq 1/2$  has the form (see, also \cite{khho})
\bb
E_{n,l} = m\frac{n+\sqrt{(l+1/2)^2-a^2}}
 {\sqrt{[n+\sqrt{(l+1/2)^2-a^2}]^2+a^2}}, \quad n = 0,1,2 \ldots , -\infty<l<\infty,
\label{spectrumw}
\ee
It can be easily shown that the discrete spectrum accumulates at the point $E=m$, and its asymptotic form as $n\gg 1$ is given by the formula
\bb
E_{n,l}=m-\frac{ma^2}{2n^2}.
\label{nonen}
\ee
In the region $1/2>\gamma>0$ the wave function $\sim U_2(r,E)/\sqrt{r}$ is singular but square-integrable at $r\to 0$ with respect to the measure $r dr$ and
the generalized eigenfunctions $F(r, E)$ of the radial Hamiltonian
should be chosen in the form
\bb
F(r,\gamma,E)=U_1(r,\gamma,E)+\xi U_2(r,\gamma,E),
\label{sol2}
\ee
where $-\infty\le\xi\le\infty$ is the real parameter. For each $l=-1,0$ there exist self-adjoint Dirac Hamiltonians $h_{\xi}$ parameterized by $\xi$ (the values $\xi=\pm \infty$ are equivalent) and specified by the asymptotic  self-adjoint boundary conditions at the origin.

For $0<\gamma<1/2$, the coefficient $B_l(\gamma,E,\xi)$ takes the form
\bb
B_l(\gamma, E,\xi)=\frac{\Gamma(\gamma+1+iaE/p)}{\Gamma(\gamma+1-iaE/p)}e^{-2i\eta(\gamma)-i\pi(|l+1/2|-\gamma)}+ \nonumber\\
+\xi\left(\frac{2p}{m}\right)^{2\gamma}\frac{\Gamma(\gamma+1+iaE/p)\Gamma(-2\gamma+1)}
{\Gamma(\gamma+1-iaE/p)\Gamma(2\gamma+1)}e^{-2i\eta(-\gamma)+i\pi(|l+1/2|+\gamma)}.
\label{coef1xi}
\ee
Performing the analytic continuation in the region $E<m$ by formula (\ref{1sheet}), we obtain
the equation that determines discrete energy levels of  bound states in the form
$B_l(\gamma, \lambda, E, \xi)=0$. For $\xi=0$, the entire analysis is similar to the one above, and all the formulas obtained there remain applicable in the case $0<\gamma<1/2$.

We fail to derive an explicit formula for the discrete energy spectrum in this region,
but we can show: 1) the discrete energy levels are in the region $-m\leq E<m$ for $\xi\neq 0$, i.e.
there exist values $\xi$ at which the lowest energy level can reach the boundary of the lower energy continuum $E=-m$, 2) the spectrum accumulates at the point $E = m$ and is described by formula (\ref{nonen}), independent of $\xi$.

For $\gamma=i\sigma$ the behavior of the two functions $\sim U_{1,2}(r, i\sigma,E)/\sqrt{r}$ essentially differ from the one for real $\gamma$: both these functions
oscillate near $r\to 0$. So, the most correct representation for the generalized eigenfunctions $F(r,\gamma,E)$ has to be
\bb
F(r,\gamma,E)=U_1(r,i\sigma,E)e^{i\theta}+ U_2(r,i\sigma,E)e^{-i\theta},
\label{sol2}
\ee
where $0\le\theta\le\pi$ is the real parameter in which case $\theta=0,\pi$ are equivalent. For $\gamma=i\sigma$, there exists  one-parameter family of self-adjoint Hamiltonians $h_{\theta}$ parameterized by the $\theta$ and specified by the asymptotic  self-adjoint boundary conditions at the origin. For  $\gamma=i\sigma$ discrete energy levels are also identified as the
poles of the scattering amplitude at the analytic continuation in the region $-m\leq E<m$ located on the first sheet of the Riemann surface or as the zeros
of the $B_l(\sigma,\lambda,E,\theta)$ in the region $-m\leq E<m$.
Having made the analytic continuation of $B_l(\sigma,E,\theta)$ in the region $-m\leq E<m$
by formula (\ref{1sheet}), we obtain
\bb
B_l(\sigma, E,\theta)=\frac{\Gamma(i\sigma+1+aE/\lambda)}{\Gamma(i\sigma+1-aE/\lambda)}e^{-2i\eta(i\sigma)-i\pi|l+1/2|}+ \nonumber\\
+e^{-2i\theta}\left(\frac{2\lambda}{m}\right)^{2i\sigma}\frac{\Gamma(i\sigma+1+aE/\lambda)\Gamma(-2i\sigma+1)}
{\Gamma(i\sigma+1-aE/\lambda)\Gamma(2i\sigma+1)}e^{-2i\eta(-i\sigma)+i\pi|l+1/2|}
\label{coefthet}
\ee
and then derive the equation determining discrete energy levels ($B_l(\sigma, \lambda, E, \theta)=0$) on the first sheet of the Riemann surface in the form
\bb
 \frac{\Gamma(2i\sigma)\Gamma(-i\sigma-aE/\lambda)}{\Gamma(-2i\sigma)\Gamma(i\sigma-aE/\lambda)}
 e^{-2i(\sigma\ln(2\lambda/m)+\pi|l+1/2|)}=-e^{2i\theta},
\label{coef1theta}
\ee
or
\bb
-\sigma\ln(2\lambda/m)+\arg\Gamma(2i\sigma)-\arg\Gamma(i\sigma-aE/\lambda)-\pi|l+1/2|-\theta =k\pi,\quad k=0,\pm 1, \ldots .
\label{enreal}
\ee

We emphasize that Eq. (\ref{enreal}) determines the fermion energy levels in the region
$m>E\geq-m$ implicitly. Analysis of Eq. (\ref{enreal}) shows  that the number of discrete energy levels is finite in the interval $-m\le E < 0$, and the spectrum for $k\gg 1$ is described by the right-hand side of asymptotic formula (\ref{nonen}). We also note that near the boundary of the lower energy continuum, i.e., at $E=-m+\epsilon,\epsilon>0$, Eq. (\ref{enreal}) closely resembles
the formula for the electron energy spectrum in a strong cutoff Coulomb potential
in 2+1 dimensions near $E=-m$ (see [19]).

One can derive an explicit formula for energy levels near the boundary of the lower energy continuum  $E=-m + \epsilon, \epsilon>0$, in the limit $\sigma\ll 1$. Using formulas
\bb
\frac{\Gamma(2i\sigma)}{\Gamma(-2i\sigma)}\approx -(1+4i\sigma\psi(1))\equiv e^{i\pi-4i\sigma{\cal C}}, \quad \frac{\Gamma(-i\sigma-aE/\lambda)}{\Gamma(i\sigma-aE/\lambda)}\approx \phantom{mmmmm}\nonumber\\
\approx 1-2i\sigma\psi\left(\sqrt{\frac{a^2m}{2\epsilon}}\right)
\equiv e^{-i\sigma\ln(a^2m/2\epsilon)+i\sigma\sqrt{2\epsilon/ma^2}},
\label{transf}
\ee
where $\psi(z)$ is the logarithmic derivative of Gamma function  \cite{GR}, ${\cal C}=-\psi(1)=0.57721$ is the Euler constant,  as well as  formula
\bb
\psi(z)|_{z\to\infty}\approx \ln z-\frac{1}{2z}-\frac{1}{12z^2},
\label{psias}
\ee
we obtain
\bb
\sigma\sqrt{\epsilon_{n,l}/2ma^2}\approx \theta-\pi/2+\sigma(\ln 2a+2{\cal C})+\pi n,\quad n=0,\pm 1, \ldots \label{grounm}
\ee
For $0<\sigma\ll 1$ Eq. (\ref{grounm}) has real solution $E=-m,\epsilon_{n,l}=0$ for $n=0$ and $\sigma_{cr}(a_{cr},l_{cr})$ as a function of $a_{cr},l_{cr}$ related to $\theta_{cr}=\pi/2-\sigma_{cr}(\ln 2a_{cr}+2{\cal C})$.
It should be noted the various values of self-adjoint extension parameter lead
to inequivalent physical cases (see, for example, \cite{phg,asp1,as0}
and a choice of definite value requires additional physical arguments
which implies that  each extension can  be understood through
an appropriate physical regularization \cite{gtv1}.
It should be emphasized that in the supercritical Coulomb potential with $\gamma=i\sigma$,
when solving the problem by the physical regularization procedure, which
is applied in the conventional quantum mechanics, the stronger singularity
of the Coulomb potential at the origin has to be regularized by a cutoff radius
$R$ of a Coulomb potential at small distances $r$. Therefore, physically,
in such a supercritical potential the self-adjoint extension parameter
can be interpreted  in terms of product $mR$.
So, the nonzero value $\theta_{cr}$ at which the lowest energy level reaches the boundary $E=-m$
obviously means that when solving the problem by the physical regularization,
the the lowest energy level can become equal $-m$  only in a supercritical (but cutoff) Coulomb potential.

\section{Fermions pair production}

{\bf Massive case}.
Now we show that if $a, \sigma$ are further increased, the lowest energy level
dives into the lower continuum ($E<-m$) and becomes a quasi-stationary state
(a resonance) at some $a>a_{cr},\sigma>\sigma_{cr}$.
Hence, as we allow a small change in $\sigma$ such that $\sigma>\sigma_{cr}$,
a sudden change in spectrum has to occur.
For which reason the scattering amplitude has a discontinuity associated with the disappearance
of its bound state pole from the physical sheet: for $E<-m$ only the continuous spectrum exists, but below ${\rm Re}E>0, {\rm Im}E>0$ there is a second (nonphysical, (${\rm Re}E<0, {\rm Im}E<0$, ${\rm Re}\sqrt{m^2-E^2}<0, {\rm Im}\sqrt{m^2-E^2}<0 $) sheet on which now the former bound state resides at $\sigma>\sigma_{cr}$.
The key difference of the case $\sigma>\sigma_{cr}$ from $\sigma<\sigma_{cr}$ is that the former bound states at $\sigma>\sigma_{cr}$ become quasi-stationary ones; they have ``complex energies'' $E=|E|e^{i\tau}$. It follows from the equation (\ref{grounm})
that the diving point ($\epsilon=0$) defines a critical coupling,
$a_{cr}$, and strongly depends on the self-adjoint extension parameter
(i.e., physically, on the cutoff radius $R$ of a Coulomb potential).

In order to make the needed analytic continuation we must determine $B_l(\sigma,E,\theta)$  on the upper edge of the cut chosen to run along the negative real energy axis from the branch point $-m$ to $-\infty$ and then have to go across the cut on the second sheet (see, for example, \cite{rtkh}. As a result, we derive the transcendental complex equation that determine implicitly the ``complex energies'' of the quasi-stationary states  in the form
\bb
 \frac{\Gamma(2i\sigma)\Gamma(-i\sigma-iaE/p)}{\Gamma(-2i\sigma)\Gamma(i\sigma-iaE/p)}
 e^{-2i(\sigma\ln(-2ip/m)+\pi|l+1/2|)}=-e^{2i\theta}.
\label{coef2theta}
\ee
For $\sigma\ll 1$ it is only natural to look for solutions of this equation in the  $E=|m + \epsilon|\exp(i\tau), 1\gg\epsilon>0$. Then, using formulas
\bb
 \frac{\Gamma(-i\sigma-iaE/p)}{\Gamma(i\sigma-iaE/p)}|_{\sigma\ll 1}
\approx e^{-2i\sigma({\rm Re}\psi(z)+{\rm Im}\psi(z)}, z=-iaE/p\equiv -ia|E|\exp(i\tau)/p,
\label{transf2}
\ee
$$
{\rm Im}\psi(ix)=\frac{1}{2x}+\frac{\pi}{2}\coth \pi x, {\rm for}\quad x\gg 1,
$$
as well as Eq. (\ref{psias}), we find
$\tau \approx \pi+(\pi/2)e^{-\sqrt{2m(\pi a)^2/\epsilon}}$ and
\bb
 \sigma\epsilon_{n,l}/6ma^2 \approx \theta-\pi/2+\pi n+\sigma(2{\cal C}+\ln 2a).
\label{boun}
\ee
We  see  that the rights of equations (\ref{grounm}) and (\ref{boun}).

It is seen that when $\sigma>\sigma_{cr}$, the lowest state dives into the
negative energy continuum and becomes a quasi-stationary state (a quasi-localized resonance)
described with the quasi-discrete  spectrum $E={\rm Re}E - i|{\rm Im}E|$.

Physically, the appearance of new fermion states with negative energies
implies a rearrangement of the vacuum. The additional distortion of the negative
energy continuum (due to the quasi-stationary states)  leads to a negative charge
density due to the ``real vacuum polarization'' \cite{grrein}. The energies of the
quasi-stationary states defines and depends upon the  parameter $\theta$.

It follows from the equation (\ref{boun}) that at $1\gg\sigma>\sigma_{cr}$
\bb
{\rm Re}E \approx -m -\epsilon_s,\quad  \epsilon_s \approx 6ma^2_{cr}\frac{\sigma-\sigma_{cr}}{\sigma_{cr}}(2{\cal C}+\ln 2a_{cr}).
\label{boun1}
\ee
The modulus of the imaginary part $E$ $w=2|{\rm Im}E|\sim me^{-\sqrt{2m(\pi a)^2/\epsilon_s}}$ is the doubled probability of the production  of the fermion pair by supercritical Coulomb field.  It is exponentially small in this case and the lifetime of the quasi-stationary supercritical level  is diverging $\Delta t\sim 1/w$.

If $a$ is further increased, other
levels will sequentially dive into the lower energy continuum  at higher $a$ (see, for instance, \cite{grrein}).

{\bf Massless case}.
In the massless case the spectrum is continuous everywhere and the  bound states are
absent, nevertheless, the quasi-stationary states emerge in a supercritical
Coulomb potential. In solving the problem with massless fermions, it is evident,
we can use the some needed formulas (putting in them $m=0$) which we have derived
above for the massive case. In particular, the main equation  (\ref{main0})
obviously is valid for the massless case if we put in it: $m=0, p=|E|, x=-2i|E|r,
E/p=e'\equiv {\rm sign} E$.

In the region $|E|>0$  the energy spectrum  is continuous for real $\gamma$ and a charged massless fermion in a Coulomb potential does not have bound states. Nevertheless, it is helpful
to analyze the neighborhood of zeros of the coefficient in the ingoing wave; they, for example,
may characterize some kind of accumulation points of fermion states.
It is rewarding to study directly the case $1/2>\gamma>0$. This case is described
by the equation (\ref{coef1xi}) with taking into account the above replacements.
Then, solving the equation $B_l(\gamma,E,\xi)=0$, we obtain  the following complex equation
\bb
\left(2\frac{|E|}{E_0}\right)^{-2\gamma}=\xi\left[\frac{\Gamma(1-2\gamma)\Gamma(\gamma-ie'a)}
{\Gamma(1+2\gamma)\Gamma(-\gamma-ie'a)}\right]e^{i\pi(e'\gamma-1/2)}
\label{bouncomp}\ee
and using the known representation
\bb
\arg\Gamma(x+iy)=y\left[-{\cal C}+\sum\limits_{n=1}^{\infty}\left(\frac1n-\frac1y\arctan\frac{y}{x+n-1}\right)\right],
\label{Garep}
\ee
we rewrite (\ref{bouncomp}) in the form of two real equations:
\bb
E(a,\gamma,\xi)=\frac{e'}{2}E_0\left[\frac{\Gamma(1+2\gamma)|\Gamma(-\gamma-ie'a)|}
{|\xi|\Gamma(1-2\gamma)|\Gamma(\gamma-ie'a)|}\right]^{1/2\gamma}
\label{boungam0}\ee
and
\bb
 \pi\left(e'\gamma-\frac12\right)+\arctan\frac{e'a}{\gamma}-
  -\sum\limits_{n=1}^{\infty}\arctan\frac{2e'a\gamma}{n^2+a^2-\gamma^2}=(s-1)\frac{\pi}{2},
\label{boundri20}\ee
where $s=\xi/|\xi|=\pm 1$, $s=1 (-1)$ for $\infty>\xi\geq 0 (0\geq \xi>-\infty)$ and
we introduce the positive constant $E_0$ with the dimension of mass. We see that
an energy scale in the massless case set explicitly by the parameter $E_0$.
The values $a, \gamma, s, e', \xi$  have to be determined by these two equations.

For  $\gamma\to 1/2$ near $|E|=0$, we find
\bb
E=e'\frac{1-2\gamma}{2|\xi|}\frac{|\Gamma(-1/2-ie'a)|}{|\Gamma(1/2-ie'a)|},
\label{newb}
\ee
whence it follows that $|E|=0$  and equality (\ref{boundri20}) is satisfied
by the value $\gamma=1/2$ for $e'=1$ and $s=1$ and for $e'=-1$ and $s=-1$ only if  $a^2=(l+1/2)^2-1/4$, i.e. only in free case $a=0$. In the free case, the quantum system exhibits the particle-antiparticle  symmetry.

For $\gamma\to 0$, $|E|$ tends to $0$  as
$2E\approx e'(1/|\xi|)^{1/2\gamma}$ and equality (\ref{boundri20})
is satisfied by $e'=\pm 1$, $\gamma=0$ only if $s=-1(0\geq\xi>-\infty)$.
This implies that particle states with $E>0$ and, on the other hand
antiparticle states with $E<0$ accumulate near the point $E=0$ if $|\xi|>1$  but remain
separated as long as $a<a_{cr}=1/2$ (see \cite{20}).

As $a>a_{c}=1/2$ the energy spectrum
of massless fermions has to be changed; the scattering amplitude has a discontinuity
associated now with the appearance of quasi-stationary states with
negative ``complex" energies ($E=|E|e^{i\beta}$),
which are located on the second sheet. It will be recalled that
only the stationary states  with  the continuous spectrum exists in the
region $E<-m$ on the physical sheet and the point $E=0$ is the branch point
of the  scattering amplitude in the complex plane of $E$.

Now we need to determine the coefficient $B_l(\sigma,E, m=0, \theta)$ given on the upper edge of the cut $(-0, -\infty)$ in going across the cut on the second sheet. As a result, we derive
the following equations ($\gamma=i\sigma$)
\bb
\frac{|\Gamma(i(e'a-\sigma))|}{|\Gamma(i(e'a+\sigma))|}
e^{-\pi\sigma+2\sigma\beta}=1, \quad e'=-1
\label{boundsig10}\ee
and
\bb
2\sigma\ln(2|E|/E_0)-2\arg\Gamma(2i\sigma)+\arg\Gamma(-ia+i\sigma)-\arg\Gamma(-ia-i\sigma)= 2(\theta -\pi n -\pi/2).
\label{bounsup}
\ee
Here $n=1,2 \ldots $, $\pi\geq \theta\geq 0$ and now $e'=-1$  corresponds to the nonphysical sheet.
Increasing $a$ ($\sigma$) will increase  $n$ and decrease the energy.
 Using the equation (\ref{Garep}) we write Eq. (\ref{bounsup}) for $\sigma\ll 1$ in the form
\bb
\sigma\ln(2|E|/E_0)=(\theta -\pi n -\pi/2) -\sigma{\cal C}-\sigma\ln\sqrt{1+a^2}.
\label{bounsmal}
\ee


For small $\sigma\ll 1$,  Eq. (\ref{boundsig10}) has approximate  solution
$\beta \approx-1/2a+\mathrm{Im}\psi(ia)+\pi/2$ and
  $\beta \approx [1+\coth(\pi/2)]\pi/2\approx (1+0.04)\pi$ for $a=1/2$.
Eqs. (\ref{boundsig10}) and (\ref{bounsmal})
are approximately satisfied near $|E|=0$ only
in the region $E<0$.  Then, for $\sigma\ll 1$ we find
the spectrum of supercritical resonances in the form
\bb
 E_{n,\beta,\sigma, \theta}= (E_0/2) \cos\beta
 e^{-(\pi n-\theta+\pi/2+\sigma {\cal C}+\sigma\ln\sqrt{1+a^2})/\sigma}.
\label{energyres}
\ee
 This spectrum, as function of $a$, has an essential singularity at
$a=a_c, \sigma_ñ=0$ and the infinite number of quasi-discrete levels
occurs \cite{as11b,ggg,gss}. In the massless case there is no natural length
scale to characterize the localization region of the infinite number
of emerging quasi-stationary  states.
The stronger singularity of the Coulomb potential at the origin has to be
regularized in the supercritical regime,  by a finite size
$R_i$ of the Coulomb impurity and the dimensionless self-adjoint extension parameter $\theta$
can now be interpreted  in terms of product $E_0R_i$.

It should be emphasized  that in the massless case there is
no sequential diving into the lower energy continuum
but the infinite number of quasi-stationary states
occurs.   It is seen that the
energy spectrum of these states is quasi-discrete, consists of a number
of broadened levels whose width (defined by the imaginary part $E$)
is related to the inverse lifetime (decay rate) of a $n$-quasi-stationary state.
These quasi-localized states  have negative energies and  are directly associated
with the positron creation in the quantum electrodynamics \cite{nkpnp}.
Again the modulus of the imaginary part $E$
\bb
w=2|{\rm Im}E|=E_0 |\sin\beta|
 e^{-(\pi n-\theta+\pi/2+\sigma {\cal C}+\sigma\ln\sqrt{1+a^2})/\sigma}
\label{wm0}
\ee
is the doubled probability of the creation  of the massless fermion pair
by supercritical Coulomb field.

\section{Summary}

In this paper we study the creation of  charged (massive and massless)
fermion pair by  supercritical Coulomb field.
We construct the self-adjoint  two-dimensional Dirac Hamiltonians with a
singular Coulomb potential and determine the quantum-mechanical states
for  self-adjoint Hamiltonians
in the corresponding Hilbert spaces of square-integrable functions.
The domain (including the singular $r = 0$ region) in these spaces,
parameterized by extension parameters and classified by boundary conditions at $r=0$,
is found for each self-adjoint Hamiltonian.

We determine the scattering amplitude as a function of the ``complex energy"
in which  the dimensionless self-adjoint extension parameter is incorporated and
then obtain the equations implicitly defining the possible discrete
energy spectra of the self-adjoint Dirac Hamiltonians with a Coulomb potential.
We establish that the quantum system in the presence of
 supercritical Coulomb potential becomes unstable which manifests
in the appearance of quasi-stationary states  in the lower (negative) energy continuum.
The above scattering amplitude  in
the presence of a supercritical  Coulomb potential is shown to become
ambiguous function; it has  a discontinuity in the complex plane of energy and
additional singularities on the negative energy axis of the nonphysical
sheet of the Riemann surface.
The imaginary part of energy  is related
to the inverse lifetime  of the quasi-stationary state as well as the creation probability
 of charged fermion pair by supercritical Coulomb field.
Explicit analytical expressions for the creation probabilities
 of charged (massive or massless) fermion pair  are obtained in a supercritical Coulomb field.


\begin{thebibliography}{55}



\bibitem{003} Y.B. Zel'dovich and V.S. Popov, Sov. Phys. Uspekhi, {\bf 14}, 673 (1972).

\bibitem{004} A. B. Migdal, {\sl Fermions and Bosons in Strong Fields} (in Russian, Nauka, Moscow,  1978).


\bibitem{blp} V.B. Berestetzkii, E.M. Lifshitz, and L.P. Pitaevskii,
{\sl Quantum Electrodynamics}, $2^{nd}$ ed. (Pergamon, New York, 1982).

\bibitem{grrein} W. Greiner, J. Reinhardt, {\sl Quantum Electrodynamics},
$4^{th}$ ed. (Springer-Verlag, Berlin Heidelberg, 2009).

\bibitem{01} E. Wichmann and N.M. Kroll, Phys. Rev. {\bf 96}, 232 (1954); Phys. Rev. {\bf 101}, 843 (1956).
\bibitem{02} L.S. Brown, R.N. Cahn, and L.D. McLerran, Phys. Rev. {\bf D12}, 581 (1975).
\bibitem{03} M. Gyulassy, Phys. Rev. Lett., {\bf 33}, 921 (1974); Nucl. Phys. {\bf A244}, 497 (1975).
\bibitem{04} A.A. Grib, S.G. Mamaev, and V.M. Mostepanenko, {\sl Vacuum Quantum Effects
 in Strong Fields}, [in Russian] (Energoatomizdat, Moscow, 1988).
\bibitem{khho}    V.R. Khalilov and C.-L. Ho, Mod. Phys. Lett. {\bf A13}, 615 (1998).
\bibitem{vp11a} V. M. Pereira, J.  Nilsson, and A.H. Castro Neto, Phys. Rev. Lett. {\bf 99}, 166802 (2007).
\bibitem{as11b} A.V. Shytov, M.I. Katsnelson, and L.S. Levitov, Phys. Rev. Lett. {\bf 99}, 236801 (2007).
\bibitem{as112} A.V. Shytov, M.I. Katsnelson, and L.S. Levitov, Phys. Rev. Lett. {\bf 99}, 246802 (2007).
\bibitem{mmf} M.M. Fogler, D.S. Novikov, and B.I. Shklovskii, Phys. Rev. {\bf B76}, 233402 (2007).

\bibitem{vmvn} V.M. Pereira, V.N. Kotov, and A.H. Castro Neto, Phys. Rev. {\bf B78}, 085101 (2008).
\bibitem{ggg} O.V. Gamayun, E.V. Gorbar, and V.P. Gusynin, Phys. Rev. {\bf B80}, 165429 (2009).
\bibitem{wzq} W. Zhu, Z. Wang, Q. Shi, K. Y. Szeto, J. Chen, and J. G.
Hou, Phys. Rev. {\bf B79}, 155430 (2009).
\bibitem{6} K.S. Novoselov et al. Science, {\bf 306}, 666 (2004).
\bibitem{7} A.H. Castro Neto, F. Guinea, N.M. Peres, K.S. Novoselov,
 and A.K. Geim, Rev. Mod. Phys., {\bf 81}, 109 (2009).
\bibitem{8} N. M. R. Peres, Rev. Modern Phys., {\bf 82}, 2673–2700 (2010).
\bibitem{review}  V. N. Kotov, B. Uchoa, V. M. Pereira, F. Guinea, and A. H. Castro
Neto, Rev. Mod. Phys., {\bf 84}, 1067-1125 (2012); arXiv:1012.3484v2 [cond-mat.str-el] (2010).
\bibitem{10} K. S. Novoselov, A. K. Geim, S. V. Morozov, D. Jiang, M. I. Katsnelson, I. V. Grigorieva, S. V. Dubonos, and A. A. Firsov, Nature, {\bf 438}, 197–200 (2005).
\bibitem{11} Z. Jiang, Y. Zhang, H. L. Stormer, and P. Kim, Phys. Rev. Lett., {\bf 99}, 106802 (2007).
\bibitem{15} A. K. Geim and K. S. Novoselov, Nature Mater., {\bf 6}, 183 –191 (2007).


\bibitem{datdc} D. Allor, T. D. Cohen, and D. A. McGady, Phys. Rev. {\bf D78}, 096009 (2008).
\bibitem{ggvo} J. Gonzarlez, F. Guinea, and M.A.H. Vozmediano, Nucl.
 Phys. {\bf B424}, 595 (1994).
\bibitem{13} Y. Zhang, Y. W. Tan, H. L. Stormer, P. Kim, Nature {\bf 438}, 201 (2005).
\bibitem{gr0} Y. Wang, V.W. Brar, A.V. Shytov, Q. Wu, W. Regan,
 H.-Z. Tsai, A. Zettl, L.S. Levitov, and M.F. Crommie, Nat. Phys. {\bf 8}, 653 (2012).
\bibitem{gr1} Y. Wang, D. Wong, A. V. Shytov, V. W. Brar, S. Choi,
 Q.Wu, H.-Z. Tsai, W. Regan, A. Zettl, R. K. Kawakami, S.G. Louie,
 L.S. Levitov, and M. F. Crommie, Science, {\bf 340}, 734 (2013).
\bibitem{gr2} A. Luican-Mayer, M. Kharitonov, G. Li, C.-P. Lu,
 I. Skachko, A.-M. B. Gon¸calves, K. Watanabe, T. Taniguchi, and E.Y. Andrei,
 Phys. Rev. Lett. {\bf 112}, 036804 (2014).

\bibitem{khim} V.R. Khalilov, I.V. Mamsurov, Mod. Phys. Lett. {\bf A31}, No 7, 1650032 (2016).
\bibitem{gkg1} F. Guinea, M.I. Katsnelson, and A.K. Geim, Nat. Phys. {\bf 6}, 30 (2009).
\bibitem{khim1} V.R. Khalilov, I.V. Mamsurov, Phys. Lett. {\bf B769}, 152 (2017).
\bibitem{sil} E. O. Silva, Eur. Phys. J. {\bf C74}, 3112 (2014).
\bibitem{27} Y. Hosotani, Phys. Lett. {\bf B319}, 332 (1993).
\bibitem{4} C.R. Hagen, Phys. Rev. Lett. {\bf 64}, 503 (1990).




\bibitem{khpr} V.R. Khalilov, Phys. Rev. {\bf A71}, 012105 (2005).
\bibitem{khle1}  V.R. Khalilov and K.-E. Lee, Journ. Phys., {\bf A44}, 205303 (2011).

\bibitem{khle0} V.R. Khalilov and K.-E. Lee, Theoretical and Mathematical Physics,  {\bf 169}(3), 1683 (2011).
\bibitem{vgt} B.L. Voronov, D.M. Gitman, and I.V. Tyutin,
Theoretical and Mathematical Physics, {\bf 150}, 34 (2007).

\bibitem{GR} I.S. Gradshteyn and I.M. Ryzhik, {\sl Table of Integrals,  Series, and Products},
$5^{th}$ ed. (Academic Press, San Diego, 1994).


\bibitem{ll} L.D. Landau and E.M. Lifshitz,
{\sl Quantum Mechanics}, 3rd ed. (Pergamon, New York, 1977).

\bibitem{phg} Ph. Gerbert, Phys. Rev. {\bf D40}, 1346 (1989).


\bibitem{asp1} F.M. Andrade, E.O. Silva, M. Pereira, Phys. Rev., {\bf D85}(4), 041701(R) (2012).


\bibitem{as0} F.M. Andrade, E.O. Silva, Phys. Lett., {\bf B719}(4-5), 467 (2013).


\bibitem{gtv1} D.M. Gitman, I.V. Tyutin, and B.L. Voronov,
{\sl Self-adjoint Extensions in Quantum Mechanics}
(Springer Science+Business Media, New York, 2012).

\bibitem{rtkh} V.N. Rodionov, I.M. Ternov, and V.R. Khalilov, ZhETF, {\bf 71}, No 9, 871  (1976).

\bibitem{20} V.R. Khalilov, Eur. Phys. J. {\bf C73}(8), 2548 (2013).

\bibitem{gss} K.S. Gupta, and S. Sen,  Mod. Phys. Lett. {\bf A24}, 99 (2009).

\bibitem{nkpnp} A.H. Castro Neto, V.N. Kotov, V.M. Pereira, J. Nilsson, N.M. Peres and B. Uchoa,
Solid State Commun., {\bf 149}, 1094 (2009).














\end{thebibliography}
\end{document}